\documentclass[prd,superscriptaddress,nofootinbib,showpacs,showkeys,preprint]{revtex4}
\usepackage{graphicx}
\usepackage[usenames,dvipsnames]{color}
\usepackage{amsmath,amssymb}
\usepackage[colorlinks]{hyperref}
\usepackage{float}

\newcommand{\be}{\begin{equation}}                                             
\newcommand{\ee}{\end{equation}}

\usepackage{epsfig}
\makeatletter

\newcommand{\Rmnum}[1]{\expandafter\@slowromancap\romannumeral #1@}
\makeatother

\newcommand{\noplus}{}

\newcommand{\tmop}[1]{\ensuremath{\operatorname{#1}}}

\begin{document}
\title{Infra-red Abelian dominance without Abelian-projection}
\author{Haresh Raval}
\email{haresh@phy.iitb.ac.in}
\affiliation{Department of Physics, Indian Institute of Technology, 
Bombay, Mumbai - 400076, India}
\author{Urjit A. Yajnik}
\email{yajnik@iitb.ac.in}
\affiliation{Department of Physics, Indian Institute of Technology, 
Bombay, Mumbai - 400076, India}

\begin{abstract}
Maximal Abelian gauge has been a particular choice to study dynamical generation of 
off-diagonal gluon masses in QCD. This gauge is a special case of Abelian projection. 
Massive off-diagonal gluons are considered a clue to Abelian dominance. Here we propose 
a gauge condition which is quadratic in fields and which does not fall in the class 
of Abelian projection. However it does generate off-diagonal gluon masses dynamically 
thus hinting at Abelian dominance in this gauge too. 
\end{abstract}
 
\maketitle

\section{Introduction}

One of the most burning questions to be  answered definitively in QCD is, what is the
physical mechanism by which quarks and gluons are confined. Classically
confinement is expressed as linear inter-quark potential. In mid 70s, a dual version 
of type-\Rmnum{2} superconductor was proposed by Nambu\cite{11}, 't Hooft\cite{12} and 
Mandelstam\cite{13}. In type-\Rmnum{2} superconductor magnetic field is trapped in the 
form of one dimensional Abrikosov vortex tubes inside the superconductor i.e., inside 
the medium of condensed electric charges\cite{14}. In the same way but dually the idea 
of these proposals for quark confinement mechanism is that the electric field due to 
quarks is trapped  in the form of vortex tubes in a phase in which magnetic monopoles 
are condensed so that the confinement forces can be characterised by constant string 
tension or as a linear inter-quark potential. Here the key point is that the 
dual picture is based on Abelian gauge theory whereas QCD is non-Abelian, therefore 
one needs to demonstrates that QCD reduces to an effective Abelian theory at infra-red 
scale. Secondly it requires a new concept of condensed magnetic monopoles. The need 
for such an  effective theory led to the concept of  ``Abelian 
dominance"\cite{15}. 

According to Abelian dominance, at low energy scale, 
 QCD can be effectively expressed in terms of Abelian gauge
degrees of freedom \cite{15}. It is usually discussed in terms of off-diagonal gluons 
i.e., gluons not associated with Cartan subalgebra of  $SU(N)$. In the  $SU(N)$ gauge 
theory there are $N(N-1)$ off-diagonal gluons. These gluons attaining a large dynamical 
mass is presumed to provide the required  evidence of existence of Abelian dominance. 
At infra-red  energies below this dynamical mass scale  these off-diagonal components 
would become  inactive and decouple, and only dominant degrees are now  diagonal gluons 
since they remain massless. Thus one gets $N-1$ copies of Abelian gauge theory at 
low energy scale corresponding to each diagonal gluon. As far as we know the occurrence
of off-diagonal gluon masses and  infra-red Abelian dominance have been studied mostly 
in maximal Abelian gauge, a few of the references being \cite{16,17,18,19}, which is  
a particular case of  Abelian projection\cite{12}. 

An Abelian projection\cite{12} is a partial gauge fixing which leaves the maximal torus 
group of a group $G$ unbroken. For $SU(N)$, the gauge condition takes the form of  
variable $X(x)$ satisfying following conditions:
\begin{itemize}
\item It takes values in lie algebra of $SU(N)$.
\item It transforms according to adjoint action
\end{itemize}
\begin{equation}
X(x) \rightarrow U(x)X(x)U(x)^{-1}
\end{equation} 
   We can perform  gauge rotation  and diagonalize $X(x)$
\begin{equation}
\widehat{X(x)} = V(x)X(x)V(x)^{-1};\hspace{1 cm}  V(x)\in SU(N)   
\end{equation}
This diagonalized $X(x)$ is invariant under $U(1)^{N-1}$, maximal torus group of   $SU(N)$. 
Hence each such variable $X(x)$  defines an Abelian projection. We propose ``a quadratic gauge'' 
which  doesn't fall in the class of Abelian projection and which results  in a dynamical
mass generation for off-diagonal gluons, 
which in turn leads to Abelian dominance.
 So to the best of our knowledge this paper presents
one substantial ingredient of the confinement without Abelian projection.

\section{a quadratic gauge and Effective Lagrangian}\label{sec1}
An effective Lagrangian of the theory is obtained through gauge fixing. Unlike the usual gauges 
here we choose a  different gauge condition, namely  constraining the operator $ A^{\mu a}A^a_{\mu}$ 
viz., `` a quadratic gauge ''. i.e,
\begin{align} \label{eq: 0 }
 F^a [ A^{\mu} ( x) ] =
A^a_{\mu} ( x) A^{\mu a} ( x) = f^a ( x) ; \  \text{  for each $a$ }
\end{align}
where $f^a(x)$ is an arbitrary function of $x$. This clearly is not an Abelian projection as 
the gauge condition doesn't take values in the Lie algebra. Furthermore, the condition of abelian 
projection is stipulated to be covariant in order to ensure survival of an abelian
component. The quadratic gauge does not meet this condition either.
This gauge condition results in the gauge fixing and ghost
contributions to the effective Lagrangian of the form
\begin{align} \label{eq1}
 \mathcal{L}_{\tmop{GF}}+ \mathcal{L}_{\tmop{ghost}} =&- \frac{1}{2 \zeta}\displaystyle\sum\limits_{a}  ( A^a_{\mu} A^{\mu a})^2  - \displaystyle\sum\limits_{a}\overline{c^a}A^{\mu a} ( D_{\mu} c)^a  
 \end{align}
Now onwards we shall drop $\displaystyle\sum$ but  summation over  $a$ will be understood  where it 
appears repeatedly, including when repeated \textit{thrice} as in the ghost terms above.
In particular,
\begin{equation}
 \mathcal{L}_{\tmop{ghost}} =  - \overline{c^a} A^{\mu a} ( D_{\mu} c)^a 
= - \overline{c^a} A^{\mu a} \partial_{\mu} c^a + g f^{a b c} 
\overline{c^a} c^c A^{\mu a} A^b_{\mu}
\end{equation}
where the summation over indices $a$, $b$ and $c$ each runs independently over $1$ to $N^2-1$.
With this understanding we write the effective Lagrangian as
 \begin{eqnarray} \label{eq: 2 }
  \mathcal{L}_{\tmop{eff}} &=&- \frac{1}{4} F^a_{\mu \nu} F^{\mu \nu a}
\noplus - \frac{1}{2 \zeta}  ( A^a_{\mu} A^{\mu a})^2 - \overline{c^a}
A^{\mu a} ( D_{\mu} c)^a 
\end{eqnarray}
where first term is Yang-Mills Lagrangian with 
$ F^a_{\mu \nu}(x) = \partial_{\mu}A^a_{\nu}(x)- \partial_{\nu}A^a_{\mu}(x)-g f^{abc} A^b_{\mu}(x)A^c_{\nu}(x)$.

\section{ Off-diagonal gluon mass generation and Abelian dominance in a quadratic gauge}
Although we do not intend to derive perturbative rules for the $S$-matrix here,
the intuitive understanding in terms of the properties of quanta that can in principle
occur in the asymptotic states is the most convenient in taking the discussion forward.
As such we now proceed to identify propagators which allows us to make the hypothesis
of ghost condensation. Then we proceed to deriving  mass terms of the effective degrees of
freedom using the ghost condensation hypothesis. 
\subsection{Mass generation due to the ghost condensation}
Gluon propagator is obtained from Yang-Mills Lagrangian in the standard form,
 \begin{equation}
  \mathcal{O}_{\mu \nu}^{- 1 a b}  ( p) = -
\frac{\delta^{a b}}{p^2}  \left( \eta_{\mu \nu} - \frac{p_{\mu} p_{\nu}}{p^2} \right)
 \end{equation}
On the other hand the ghosts do not possess any propagator in the theory since 
there is no free quadratic part of ghost fields in the effective Lagrangian in eq.~\eqref{eq: 2 }. 
Hence for the propagator of a ghost field we have
\begin{equation}
 G^{a b}(p)= 0
\end{equation}
However the ghost term in eq.~\eqref{eq: 2 } does contain ghost-gluon interaction terms.  
Thus while the ghosts are non-propagating spectators, they continue to interact 
through the gluons. Consider therefore, the individual terms in the ghost Lagrangian $\mathcal{L}_{\tmop{ghost}}$
\begin{equation}\label{eq: 4 }
- \overline{c^a}A^{\mu a} ( D_{\mu} c)^a = - \overline{c^a} A^{\mu a} \partial_{\mu} c^a + g f^{a b c}
\overline{c^a} c^c A^{\mu a} A^b_{\mu}
\end{equation}
where the summation over the indices goes as explained in sec.~\ref{sec1}.  We see that if the ghosts 
undergo condensation, the gluons acquire a mass matrix. Such a possibility was elaborated in
Ref. \cite{16}.
We shall here show, that a similar condensation mechanism is responsible for the generation 
of off-diagonal gluon mass  in the quadratic  gauge. We assume that ghost fields acquire 
non-zero  real  ghost-anti{\small -}ghost condensation i.e., 
\begin{equation}
 \langle\overline{c^i} c^j\rangle  \neq 0 \hspace{10mm} \mathrm{for\  all}\quad i\quad \mathrm{and}\quad j. 
\end{equation} 
This may be understood as a consequence of the interaction among the spectator ghost fields
mediated by the gluons.
In the proposed ghost condensed phase, second term of eq.~\eqref{eq: 4 } gives us 
off-diagonal components of gluon mass matrix 
\begin{equation}
 (M^{ 2})^{a b}_{\tmop{dyn}} = 2 g \displaystyle\sum\limits_{c=1}^{N^2-1}f^{a b c} \langle\overline{c^a} c^c\rangle
\end{equation}
whereas diagonal components of $M^{ 2}_{\tmop{dyn}}$ are zero since $f^{a a c} = 0$. 
To obtain a spectrum of the theory i.e., to obtain masses of gluons, we
must diagonalize the matrix and find eigenvalues.

The required demonstration is simple in an $SU(N)$ symmetric state, where ghost-anti{\small -}ghost 
condensates are taken to be identical i.e., 
\begin{equation}
\langle\overline{c^1} c^1\rangle = ... =  \langle\overline{c^1} c^{N^2-1}\rangle = ... =  \langle\overline{c^{N^2-1}} c^1\rangle = ... =     
\langle\overline{c^{N^2-1}} c^{N^2-1}\rangle = K
\end{equation}
Thus
\begin{equation}
(M^{ 2})^{a b}_{\tmop{dyn}} = 2 g \displaystyle\sum\limits_{c=1}^{N^2-1}f^{a b c} K
\end{equation}
Hence now this gluon mass matrix is just an antisymmetric matrix formed by structure constants 
of $SU(N)$,  
\begin{equation}
J^{a b}=\Big[\displaystyle\sum\limits_{c=1}^{N^2-1} f^{a b c}\Big]
\end{equation} 
where $a$ is a row index and $b$ is a column index. What we claim is, this matrix has 
exactly $N(N-1)$ non-zero eigenvalues and thus nullity  is exactly $N-1$.

Mathematics of this result is the following: structure constants themselves 
furnish $(N^2-1) \times (N^2-1)$  adjoint representation. So,  
$J^{a b}$ is equal to sum over 
generators in adjoint representation i.e., 
$i\Big(\displaystyle\sum\limits_{c=1}^{N^2-1}[T^{a b}]^c\Big) \in \mathfrak{g}, lie\hspace{0.1 cm} algebra\hspace{0.1 cm} vector\hspace{0.1 cm} space$. 
Coadjoint representation $[T_{a b}]_c^* $ which is dual of the adjoint  is the 
same as the adjoint  for $SU(N)$. 
Elements $g$ of $SU(N)$ acts on $  \mathfrak{g^*}, dual\hspace{0.1 cm} vector\hspace{0.1 cm} space$ by 
conjugation $$\{Ad^*F = g^{-1} F g, \hspace{0.2 cm} F \in \mathfrak{g^*}\}$$ 
The  orbit $\mathcal{O}_F = $ \{${Ad^*F, \hspace{0.2 cm} \forall g \in SU(N)} $\}, 
passing through $F$, is known as coadjoint orbit. Now, any point $ F \in \mathfrak{g^*}$ 
in compact connected lie group has maximal torus group as a stabilizer. Hence for $SU(N)$ 
maximal torus group $U(1)^{N-1}$ is also a stabilizer. So the coadjoint orbits of $SU(N)$ 
group are isomorphic to $SU(N)/U(1)^{N-1}$ i.e., 
$\mathcal{O}_F \sim SU(N)/U(1)^{N-1} \sim \mathbb{C}P^{N-1}\otimes\mathbb{C}P^{N-2}\otimes...\otimes\mathbb{C}P^1$. 
So,  it is  $N(N-1)$ dimensional symplectic manifold with symplectic form $Tr(F.[T^a,T^b])$, 
where $F \in \mathfrak{g^*}$ and $T^a, T^b \in \mathfrak{g}$, defined on it.\cite{20}. 
In the present problem, the given matrix 
$J^{a b} = \frac{-i}{N}Tr(F.[T^a,T^b])$ 
with $F=\displaystyle\sum\limits_{c=1}^{N^2-1}T^c$ and $T^a,T^b,T^c$  being basis generators. 
The rank of a symplectic form is always equal to  the dimension of a coadjoint orbit.   
Since given matrix is normal, its rank and number non-zero eigenvalues are equal. So the 
rank and therefore the  number of non-zero eigenvalues are $N(N-1)$. Hence nullity is $N-1$. Thus we have proved that in an $SU(N)$ invariant vacuum,  $N(N-1)$ off-diagonal gluons acquire masses and $N-1$ diagonal gluons remain massless.

The non-zero eigenvalues thus identified, being eigenvalues of antisymmetric matrix, 
are  pure imaginary and occur in conjugate pairs. viz., 
$M^2_{gluon}=\pm im^2 \  ( m^2  \ \text{positive real})$. This implies, mass of these gluons 
$M_{gluon} = \frac{1}{\sqrt{2}}(1\pm i)m \ \text{or} \ \frac{1}{\sqrt{2}}(-1\mp i)m $. We ignore 
the latter one since it gives $ Re(M_{gluon})$ negative which is not physical. Because we are not 
interested in an $S$-matrix interpretation for these degrees of freedom, \textit{prima facie} there is
no danger from these eigenvalues being pure imaginary. However to retain the intuitive appeal 
of the arguments it is necessary to check that we have not departed too far from their 
interpretation as quanta and in principle an $S$-matrix interpretation.
This is what we shall do in the next section. 

We now emphasize the implication of the fact that the off-diagonal  gluons do have real part to 
their mass. Since they are massive they can not mediate long range interaction beyond the range 
of inverse of its mass, $ r \lesssim M^{-1}$.  Therefore below mass scale of lightest off-diagonal 
gluon $M^{\tmop{lig}}_{gluon}$, no off-diagonal gluons  will contribute to any interaction.
So, diagonal gluons are the only degrees of freedom that will mediate long range 
interactions  as they are massless. This way one ends up getting an effective description of 
theory with $N-1$ copies of Abelian gauge theory. Thus if this description as commonly used in the literature 
can be relied on, then we have proved existence of Abelain dominance at infra-red scale in this gauge.

\section{hermiticity of the effective Lagrangian in a quadratic gauge}
In order to retain the appeal to a picture of this ground state in terms of quanta it is
useful to check that it will not conflict with nominally expected restrictions on the manner
which such degrees of freedom enter into an $S$-matrix. Here we will show that while in the normal phase
the Lagrangian is manifestly hermitian, a condition  indispensable for $S$-matrix unitarity\cite{21}, the effective 
Lagrangian obeys an extended  hermiticity condition also in the ghost phase, 

The effective Lagrangian in the normal phase is  given in eq.~(\ref{eq: 2 })
\begin{eqnarray} 
  \mathcal{L}_{\tmop{eff}} &=&- \frac{1}{4} F^a_{\mu \nu} F^{\mu \nu a}
\noplus - \frac{1}{2 \zeta}  ( A^a_{\mu} A^{\mu a})^2 - \overline{c^a}
A^{\mu a} ( D_{\mu} c)^a 
\end{eqnarray}
The hermiticity property of fields is given by
\begin{eqnarray}
A_{\mu}^{a \dagger}&=& A_{\mu}^{a} \nonumber \\
c^{a \dagger}&=&c^a \nonumber \\
\overline{c^{a}}^ {\dagger}&=&-\overline{c^{a}}
\end{eqnarray}
It is easy to check that this Lagrangian is hermitian under hermitian conjugation of fields since
\begin{center}
$(\overline{c^a} c^c)^{\dagger}=-c^{c }\overline{c^{a}}= \overline{c^a} c^c $
 \ and\\
$ (\overline{c^a} A^{\mu a} \partial_{\mu} c^a)^{\dagger}=- \partial_{\mu} c^a  A^{\mu a}\overline{c^a}=\overline{c^a} A^{\mu a} \partial_{\mu} c^a $
\end{center}
(we have used anti commutativity of ghost fields)
 
In the  ghost condensed phase, the effective Lagrangian now becomes
 \begin{eqnarray} \label{5}
  \mathcal{L}_{\tmop{eff}} &=&- \frac{1}{4} F^a_{\mu \nu} F^{\mu \nu a}
\noplus - \frac{1}{2 \zeta}  ( A^a_{\mu} A^{\mu a})^2 - \overline{c^a}
A^{\mu a} ( \partial_{\mu} c)^a + M^2_a A^a_{\mu} A^{\mu a}
\end{eqnarray}
Here $M^2_a=0$ when $a$ indexes the diagonal gluons, 
e.g, for $SU(3)$, $M^2_3=M^2_8=0$. While for the off-diagonal gluons, $M^2_1=+im^2_1, M^2_2=-im^2_1, \  
M^2_4=+im^2_2,  M^2_5=-im^2_2, \  M^2_6=+im^2_3, M^2_7=-im^2_3 \ (m_1^2, m_2^2, m_3^2 $ 
are positive real$)$, hence for $SU(3)$ last term of the effective Lagrangian in eq.~\eqref{5} would be
 \begin{eqnarray} \label{6}
  M^2_a A^a_{\mu} A^{\mu a}= & +&im^2_1 A^1_{\mu} A^{\mu 1}-im^2_1A^2_{\mu}A^{\mu 2} +im^2_2A^4_{\mu} A^{\mu 4}-im^2_2A^5_{\mu} A^{\mu 5} \nonumber \\&+&im^2_3A^6_{\mu} A^{\mu 6}-im^2_3A^7_{\mu} A^{\mu 7}
\end{eqnarray}
Now taking hermitian conjugate of the Lagrangian in eq.~\eqref{5} would alter nothing in first two terms and hermiticity of the third has been proven above but, it will interchange the sign of mass term between ``conjugate" gluons. e.g,  in eq.~\eqref{6} 
 \begin{eqnarray} \label{7}
 ( M^2_a A^a_{\mu} A^{\mu a})^\dagger = & -&im^2_1 A^1_{\mu} A^{\mu 1}+im^2_1A^2_{\mu}A^{\mu 2} -im^2_2A^4_{\mu} A^{\mu 4}+im^2_2A^5_{\mu} A^{\mu 5} \nonumber \\&-&im^2_3A^6_{\mu} A^{\mu 6}+im^2_3A^7_{\mu} A^{\mu 7}
\end{eqnarray}
We now invoke triality\cite{GRS}, a property of all known hadrons, which keeps them color singlet. 
In the condensed state we require any observable excitation above this state to possess zero triality. 
Define the triality operator $\mathfrak{T}$  as
\begin{eqnarray}
\mathfrak{T}H_1\mathfrak{T}^{\dagger}=H_2 \hspace{1 cm} \mathfrak{T}H_4\mathfrak{T}^{\dagger}=H_5 \hspace{1 cm} \mathfrak{T}H_6\mathfrak{T}^{\dagger}=H_7 \hspace{1 cm} \mathfrak{T}H_3\mathfrak{T}^{\dagger}=H_8  \nonumber \\
\mathfrak{T}H_2\mathfrak{T}^{\dagger}=H_1\hspace{1 cm} \mathfrak{T}H_5\mathfrak{T}^{\dagger}=H_4\hspace{1 cm} \mathfrak{T}H_7\mathfrak{T}^{\dagger}=H_6 \hspace{1 cm} \mathfrak{T}H_8\mathfrak{T}^{\dagger}=H_3
\end{eqnarray}
Where \ $H_i$ can be $- \frac{1}{4} F^i_{\mu \nu} F^{\mu \nu i},
 \  - \frac{1}{2 \zeta}  ( A^i_{\mu} A^{\mu i})^2, \   -\overline{c^i}
A^{\mu i} ( \partial_{\mu} c)^i, \   im^2 A^i_{\mu} A^{\mu i}
 $.
It can be seen that the triality operation does not cause any 
 change in first three terms in eq.~\eqref{5} and further when operated on eq.~\eqref{7} it can be easily verified that
\begin{equation}
\mathfrak{T}( M^2_a A^a_{\mu} A^{\mu a})^\dagger\mathfrak{T}^{\dagger}= M^2_a A^a_{\mu} A^{\mu a}
\end{equation}
Thus the vacuum may be considered to furnish a non-trivial representation of triality. 
So as a result $ \mathcal{L}_{\tmop{eff}}$ in eq.~\eqref{5} is invariant under hermitian conjugation followed by triality
\begin{equation}
\mathfrak{T}\mathcal{L}_{\tmop{eff}}^{\dagger}\mathfrak{T}^{\dagger}=\mathcal{L}_{\tmop{eff}}
\end{equation}
Which replaces  usual unitarity condition  for $S$- matrix,
$S^{\dagger}S=SS^{\dagger}=1$  by $\mathfrak{T}S^{\dagger}\mathfrak{T}^{\dagger}S=S\mathfrak{T}S^{\dagger}\mathfrak{T}^{\dagger}=1$ in the ghost condensed phase.
We may understand the inclusion of triality symmetry in ensuring the unitarity of the 
$S$-matrix to be a refinement over the usual discrete internal symmetry $C$, the charge 
conjugation symmetry.

\section{Conclusion}
We have proposed the set of gauge conditions $A^a_{\mu} A^{\mu a}=f^a(x)$, for each $a$, 
and shown that in a vacuum characterised by ghost-anti{\small-}ghost condensation which 
respects $SU(N)$ invariance, 
the off-diagonal gluons acquire masses dynamically. This result strongly 
supports the  existence of Abelian dominance  at infra-red energies in quadratic gauge in QCD. 
We also proved that in the quadratic gauge, the effective Lagrangian in  the absence of 
condensation is manifestly hermitian,  giving us usual  unitarity  condition for the 
$S$-matrix  whereas in  the ghost  condensed  phase it satisfies extended hermiticity 
condition, giving  a modified unitarity condition for the $S$-matrix .

\section*{ Acknowledgements }

We would sincerely like to thank  Prof. S. Umasankar and Prof. A. Ranjan   for  very fruitful discussions  and comments. We would also like to appreciate the help of
 Mr. Sreeraj Nair.


\begin{thebibliography}{12}
\bibitem{11} Y. Nambu, \textsl{Phys. Rev. D} \textbf{10}, 4262 (1974)
\bibitem{12} G. ’t Hooft, \textsl{Nucl. Phys. B} \textbf{190}, 455 (1981) 
\bibitem{13}S. Mandelstam, \textsl{Phys. Rep.} \textbf{23}, 245 (1976)
\bibitem{14}A.A. Abrikosov, Fundamentals of the Theory of Metals (North-Holland, Amsterdam, 1988) 
\bibitem{15} Z.F Ezawa and A. Iwazaki, \textsl{Phys. Rev. D} \textbf{25}, 2681 (1982)
\bibitem{16} Kei-Ichi Kondo and Toru Shinohara, \textsl{phys. Lett. B}  \textbf{491}, 263-274 (2000) 
\bibitem{17} Kazuhisa Amemiya and Hideo Suganuma, \textsl{Phys. Rev. D} \textbf{60}, 114509 (1999)\\
Shinya Gongyo, Takumi Iritani, Hideo Suganuma, \textsl{Phys. Rev. D} \textbf{86}, 094018  (2012)

\bibitem{18} S. Hioki , S. Kitahara , S. Kiura , y. Matsubara , O. Miyamura , S. Ohno  and T. Suzuki, \textsl{Phys. Lett. B} \textbf{272}, 326 (1991)
\bibitem{19} D. Dudal, J. A. Gracey, V. E. R. Lemes, M. S. Sarandy, R. F. Sobreiro, S. P. Sorella, and H. Verschelde, \textsl{Phys. Rev. D} \textbf{70}, 114038 (2004)
\bibitem{20} A.A. Kirillov, Lectures on the orbit method, Graduate Studies in Mathematics (American Mathematical Society, Providence, RI, 2004) vol. 64  
\bibitem{21} T. Kugo, I. Ojima, \textsl{pro. theor. phys. supplement} \textbf{66} (1979)
\bibitem{GRS} M. Gell-Mann, P. Ramond and R. Slansky, \textsl{Rev. Mod. Phys.} \textbf{50} 721 (1978)  
\end{thebibliography}
\end{document}